\documentclass{INTERSPEECH2023}

\usepackage{xcolor}
  
\usepackage{bbding}
\usepackage{hyperref}
\usepackage{multirow}
\usepackage{diagbox}
\usepackage{bm}

\interspeechcameraready

\title{Rethinking the visual cues in audio-visual speaker extraction}
\name{Junjie Li$^1$, Meng Ge$^{2,4,*}$, Zexu pan$^{3}$, Rui Cao$^{1}$, Longbiao Wang$^{1,*}$, Jianwu Dang$^{1}$, Shiliang Zhang 
\thanks{$^*$ Corresponding author.}}
\address{
  $^1$ Tianjin Key Laboratory of Cognitive Computing and Application,\\ College of Intelligence and Computing, Tianjin University, Tianjin, China\\
  $^2$ Department of Electrical and Computer Engineering, National University of Singapore, Singapore\\
  $^3$ Institute of Data Science, National University of Singapore, Singapore\\
  $^4$ Shenzhen Research Institute of Big Data, Shenzhen, China}
\email{mrjunjieli@tju.edu.cn,gemeng@nus.edu.sg,longbiao\_wang@tju.edu.cn}

\begin{document}

\maketitle
 
\begin{abstract}
The Audio-Visual Speaker Extraction (AVSE) algorithm employs parallel video recording to leverage two visual cues, namely speaker identity and synchronization, to enhance performance compared to audio-only algorithms. However, the visual front-end in AVSE is often derived from a pre-trained model or end-to-end trained, making it unclear which visual cue contributes more to the speaker extraction performance. This raises the question of how to better utilize visual cues. To address this issue, we propose two training strategies that decouple the learning of the two visual cues. Our experimental results demonstrate that both visual cues are useful, with the synchronization cue having a higher impact. We introduce a more explainable model, the Decoupled Audio-Visual Speaker Extraction (DAVSE) model, which leverages both visual cues.
\end{abstract}

\section{Introduction}

\label{sec:intro}
Speech is not only the most natural way of communication between humans, but also plays an indispensable role in human-computer interaction. Unfortunately, the speech of interest is always interfered by background noise and other speakers in the real world. 
While humans have the intrinsic ability to attend to the target speaker while ignoring other interference, also known as the cocktail party problem \cite{cherry1954some}, machines have not been constructed to reach human standards.

The goal of speaker extraction is to separate target speech by filtering out environmental noise signals and other speakers' speech signals. It plays a critical role in speech pre-processing to facilitate downstream tasks, such as active speaker detection~\cite{tao2021someone}, speaker localization~\cite{qian2021multi}, speaker emotion analysis~\cite{pan2020multi}, and automatic speech recognition \cite{narayanan2014investigation,pan2022hybrid}. In recent years, tremendous efforts have been made to improve the quality of separated speech, including techniques such as permutation invariant training \cite{yu2017permutation}, Conv-TasNet\cite{luo2019conv}, dual-path RNN \cite{luo2020dual},  SpEx+ \cite{ge20_interspeech}, SpEx++ \cite{9413359}.

Human speech perception is essentially a multi-modal process. People not only listen to speech but also observe facial expressions and lip movements. According to neuroscience studies \cite{golumbic2013visual}, visual inputs enhance people's ability to focus on the speaker of interest and reduce perceptual ambiguity in noisy environments. To mimic human perceptual processes, visual cues have been widely leveraged in recent studies \cite{ pan2022speaker, lee2021looking, li2022audio, liu2023limuse}, which utilize visual cues as auxiliary information to extract corresponding target speech. Previous studies have reported great performance compared to audio-only speech separation \cite{afouras2018deep, li22ba_interspeech, gao2021visualvoice, usev21, pan2022seg}, especially in noisy environments \cite{wang2020robust}, attributed to the robustness of visual cues against acoustic noise.

There are two types of visual cues that are useful for speaker extraction: the speaker identity cue and the synchronization cue. The speaker identity cue can be learned from a single image ~\cite{chung2020facefilter,qu2020multimodal} or a video recording based on the studies of face-voice correlation. The synchronization cue is learned from a parallel video recording which contains speech-lip synchronization ~\cite{pan2021reentry} of viseme-phoneme correlation  ~\cite{wu2019time,pan2021muse} 
 information. Aldeneh et al. \cite{aldeneh2021role} have demonstrated that the performance varies depending on the articulation, indicating that the synchronization cue provides performance improvement. Another work \cite{elminshawi2022new} argues that auxiliary information is only beneficial for selecting the speaker of interest, indicating the effect of the speaker identity cue. Wang et al. \cite{wang22s_interspeech} improve the performance by introducing auxiliary loss functions to model phonetic correlation between lip motion and phoneme, and speaker-identity correlation between timbre and facial attributes.

The state-of-the-art AVSE models usually employ a visual front-end to learn the visual cues. The visual front-end is either taken from part of a pre-trained network to extract low-level visual features or is trained end-to-end to optimize speaker extraction. Such training implicitly makes use of both the speaker identity and synchronization features. However, it is unclear which, or how much information is learned from each visual cue. Therefore, how to utilize visual cues remains an open question. We believe there is still room for improvement if we could explicitly decouple the learning of the two visual cues in one speaker extraction model which is the focus of our paper.

\begin{figure*}[ht] 
\centering 
\includegraphics[width=\textwidth]{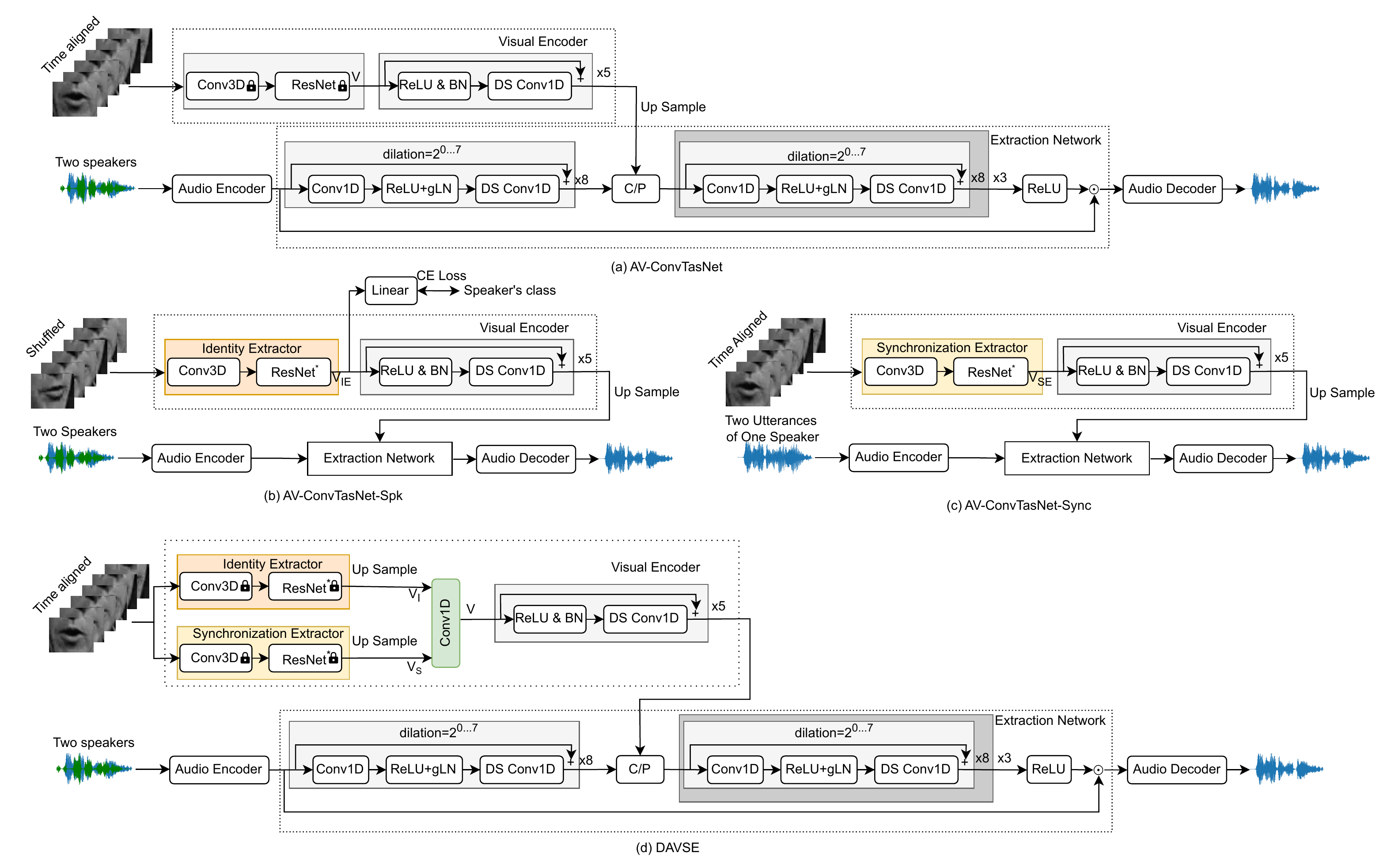}
\caption{Audio-visual speaker extraction models.
(a) AV-ConvTasNet: Raw visual streams are used to extract target speech; hence, the speaker identity cue and synchronization cue are utilized implicitly.
(b) AV-ConvTasNet-Spk: Only the speaker identity cue is utilized.
(c) AV-ConvTasNet-Sync: Only the synchronization cue is utilized.
(d) DAVSE: Both the speaker identity cue and synchronization cue are utilized explicitly. 
$\odot$ denotes point-wise multiplication, `C/P' concatenates two input embeddings over the channel dimension and projects it to a lower dimension feature using Conv1D. The module with a lock symbol denotes that its weight is fixed during training. }
\label{fig:model}
\end{figure*}

Different from~\cite{wang22s_interspeech}, we propose two different training strategies to decouple the learning of two visual cues,  namely the same-speaker aligned-visual training that is specialized in learning synchronization cue, and the different-speaker shuffled-visual training that is specialized in learning the speaker identity cue. Experimental results verify that both visual cues are useful, while the synchronization cue is clearly better. We also propose a Decoupled Audio-Visual Speaker Extraction model (DAVSE) to take advantage of both decoupled visual cues in one speaker extraction model. Our DAVSE outperforms baselines in terms of signal quality and perceptual evaluations. Our work provides a new sight into understanding the role of visual cues and presents views on how to improve the performance of AVSE.


\section{Decoupled Audio-Visual Speaker Extraction Model}
\label{sec:model}

There are two types of visual cues that can be useful for speaker extraction: speaker identity cues and synchronization cues. These cues can help identify specific speakers and extract their voices from mixed audio. Previous works typically concatenate audio and visual modalities in one model and train it end-to-end, thereby implicitly utilizing both cues. In this section, we propose two training strategies to decouple the speaker identity cue and synchronization cue from raw visual streams. Additionally, we design a Decoupled Audio-Visual Speaker Extraction model (DAVSE) that explicitly exploits both visual cues to improve speaker extraction performance.

\subsection{Typical audio-visual speaker extraction model}
A typical time-domain audio-visual speaker extraction model is exemplified by the AV-ConvTasNet \cite{wu2019time}, which contains four parts: a visual encoder, an audio encoder, an extraction network, and an audio decoder, as depicted in Figure~\ref{fig:model}(a). This model serves as our AV baseline for comparison.

The visual encoder has a 3D convolution (Conv3D) and a ResNet block followed by a video temporal convolutional block consisting 5 residual connected rectified linear unit (ReLU), batch normalization (BN) and depth-wise separable convolutional layers (DS-Conv1D) \cite{wu2019time}. The weights of Conv3D and ResNet are pre-trained according to lip reading task, similar to the work \cite{afouras2018deep}. The dimension of the output of ResNet \bm{$V$} is 512. 

The detailed architecture of audio encoder, audio decoder and extraction network can be found in work \cite{wu2019time}.

\subsection{Decoupled training for speaker identity cue}

To exploit speaker identity cue solely, we propose a different-speaker shuffled-visual training strategy, and name the model trained with this strategy as AV-ConvTasNet-Spk. The structure of AV-ConvTasNet-Spk is  the same as AV-ConvTasNet except for the visual encoder. The identity extractor has a Conv3D and a ResNet$^*$ block. The dimension of output of ResNet$^*$ \bm{$V_{IE}$} is only 256 here, as depicted in Fig.~\ref{fig:model} (b).

To extract speaker identity feature, the cross-entropy (CE) loss is added for speaker classification. According to our experience,  if only using separation loss, scale-invariant signal-to-noise ratio (SI-SNR) loss here, the model can not find a way to optimize. The training progress can be divided into two steps:

\textbf{Step 1:} The modules, Conv3D, ResNet$^*$ and Linear, are trained using CE loss. It is defined as : 
\begin{equation}
    \mathcal{L}_{CE} = -\sum_{l=0}^{L-1}\sum_{c=0}^{C-1}y_clog(softmax(\bm{W}V_{IE_{l}}))
\end{equation}
where $C$ is the number of speakers in the training dataset. $y_c$ is target speaker's class label. $W$ is a learnable weight matrix for speaker classification. $V_{IE} \in R^{L \times N}$ is output feature of identity extractor. $L$ and $N$ are time and channel dimension, respectively. 
Therefore, the identity extractor can distinguish different speakers. 

\textbf{Step 2:} We use the pre-trained identity extractor and fix these weights, and train other modules using speaker extraction loss $\mathcal{L}_{SI-SNR}$. During training, the model takes speech mixed from two speakers and shuffled visual streams of target speaker. Because of the shuffled visual streams and pre-trained identity extractor, it forbids model to learn any synchronization cue, thus solely learning the speaker identity cue to distinguish different speakers.

\subsection{Decoupled training for synchronization cue}

To exploit synchronization cue solely, we propose a same-speaker aligned-visual training strategy, and name the model trained with this strategy as AV-ConvTasNet-Sync. It shares the same structure as AV-ConvTasNet-Spk except that it doesn't have speaker classification part. The dimension of output of ResNet$^*$ \bm{$V_{SE}$} is only 256 here, as depicted in Fig.~\ref{fig:model} (c). 

During training, AV-ConvTasNet-Sync accepts speech mixed from different utterances of one speaker and time-aligned video and audio streams of the target speaker. The use of same-speaker speech mixture prevents the model from learning any identity cues, thereby allowing it to extract only the synchronization cue to extract the target speech.

\setlength\tabcolsep{4pt} 
\renewcommand{\arraystretch}{2}
\begin{table*}[ht]
	\centering
	\fontsize{8}{7}\selectfont
	\caption{SI-SNR (dB) and PESQ in a comparative study under different simulated datasets. `D-S A-V' denotes different-speaker aligned-visual dataset. `D-S S-V' denotes different-speaker shuffled-visual dataset. `S-S A-V' denotes same-speaker aligned-visual dataset. `Diff.' and `Same' denote different and same gender mixtures, respectively.  }
	\label{tab:baseline}
	\begin{tabular}{|c|c|c|c|c|c|c|c|c|c|c|c|c|c|c|c|c|c|c|c|}
		\hline
		\multirow{3}{*}{Methods}& \multicolumn{2}{c|}{\multirow{2}{*}{\#Param}}& \multirow{3}{*}{Visual Input}& 
            \multicolumn{6}{c|}{D-S A-V}& \multicolumn{2}{c|}{D-S S-V}& \multicolumn{2}{c|}{S-S A-V}\cr \cline{5-14}
		
            &\multicolumn{2}{c|}{} & &  \multicolumn{3}{c|}{SI-SNR}&\multicolumn{3}{c|}{PESQ}& SI-SNR & PESQ& SI-SNR & PESQ\cr\cline{2-3} \cline{5-14}
		& Total&Trainable& &Diff.&Same&Avg.&Diff.&Same&Avg.& Avg. & Avg. & Avg. & Avg. \cr
		\hline
		\hline
		Mixture &\diagbox{}{}  & \diagbox{}{} &\diagbox{}{} &-0.59 &-0.06 &-0.32 & 1.67& 1.73&1.70&-0.32& 1.70& 0.24& 1.79 \cr\hline

		AV-ConvTasNet \cite{wu2019time} & 16.99 M & 5.8 M&lip  & 11.32 &10.95 & 11.13 & 2.75&2.74 &2.74&-5.82&1.52&8.97&2.57 \cr
		\multirow{2}{*}{AV-ConvTasNet-Sync} & \multirow{2}{*}{9.97 M} & \multirow{2}{*}{9.97 M}  & lip&9.79 &10.45 &10.13& 2.59& 2.69& 2.64& -5.25&1.38&11.15&2.78\cr
            &  & &face &10.55 & 10.93&10.74& 2.68& 2.75& 2.72 & -4.96 & 1.41& \textbf{11.82} & \textbf{2.86}  \cr
		\multirow{2}{*}{AV-ConvTasNet-Spk}& \multirow{2}{*}{9.97 M} & \multirow{2}{*}{9.97 M}  & lip  &2.92 &-0.80 & 1.03&1.88 & 1.62& 1.75  &0.86 & 1.73  & -0.54 & 1.71\cr
            &  & & face&5.29 & -1.63&1.77 &2.16 &1.65 & 1.90&\textbf{1.51}& \textbf{1.88} &-2.28 & 1.57 \cr

            \multirow{2}{*}{DAVSE}& \multirow{2}{*}{15.32 M} & \multirow{2}{*}{4.88 M}  & lip  &12.05 &12.12 & 12.08&2.84 &2.87 &2.85 &-4.13 &1.65 & 10.73&2.72 \cr
            &  & & face& \textbf{12.77}& \textbf{12.70}& \textbf{12.73} & \textbf{2.93}&\textbf{2.95} &\textbf{2.94}&-4.41&1.63 &10.58&2.76   \cr
		 
		\hline
	\end{tabular} 
\end{table*}

\subsection{DAVSE}
To utilize both visual cues, we propose DAVSE
\footnote{\href{https://github.com/mrjunjieli/DAVSE}{https://github.com/mrjunjieli/DAVSE}}.
Unlike AV-ConvTasNet having a single branch to model visual streams and exploit speaker identity and synchronization cues implicitly. We design two branches, identity extractor and synchronization extractor in a visual encoder, to extract speaker identity feature and synchronization feature. Identity extractor and synchronization extractor are fixed, and pre-trained from AV-ConvTasNet-Spk and AV-ConvTasNet-Sync, respectively, as depicted in Fig.~\ref{fig:model} (d).

During training, DAVSE takes speech mixed from two speakers and time-aligned visual streams of target speaker. The outputs of identity extractor and synchronization extractor are concatenated along channel dimension, and then processed by 1D convolution to reduce dimension: 
\begin{align}
    \bm{V_{IS}} &= Concat(\bm{V_I},\bm{V_S}) \\
    \bm{V} &= Conv1D(\bm{V_{IS}},1,1)
\end{align}
The kernel size and stride are both set to 1. The channel dimension of \bm{$V_{IS}$} and \bm{$V$ }are 512 and 256, respectively.

\subsection{Loss function for speaker extraction}
All models are trained using scale-invariant signal to noise ratio (SI-SNR) \cite{le2019sdr}, which is defined as follows:

\begin{equation}
    \left\{ \begin{array}{lr}
    s_{target} = \frac{\hat{s}^\mathrm{T}s}{\Vert s \Vert ^2}s & \\
    e_{noise} = \hat{s} -s_{target} & \\
    SI\text{-}SNR(s, \hat{s}) =  10log_{10} \frac{\Vert s_{target}\Vert^2 }{\Vert e_{noise} \Vert^2} &\\
    \mathcal{L}_{SI\text{-}SNR}(s,\hat{s}) = -SI\text{-}SNR(s,\hat{s})
    \end{array}
    \right.
\end{equation}
where $s$ and $\hat{s}$ denote the  target speech and  estimated speech, respectively

\section{Experiments}
\label{sec:exp}
\subsection{Lip Reading Sentences 3 (LRS3) dataset}
 LRS3 \cite{afouras2018lrs3} is a large-scale audio-visual dataset that is obtained from TED and TEDx talks. There are 118,516 (408 h), 31,982 (30 h) and 1,321 (0.85 h) utterances in training, development (dev) and test sets, respectively. There are 5089 speakers in training set. The speakers in the train set and test set do not overlap. 

\subsection{Data preparation}
The audio is sampled at 16k Hz, and corresponding video frames are sampled at 25 FPS. We use face recognition \footnote{https://pypi.org/project/face-recognition/} algorithm to detect face for each frame and crop lip region from its face landmarks. Both face and lip images are used as visual input for speaker extraction task and resized to 112 * 112 pixels in greyscale.


To save computation resource, we pick 1,500 speakers and 1,000 speakers from training and dev sets, respectively.  Among each speaker, short utterances (less than 4s) are dropped and long utterances are cut to 4$\sim$6s randomly. And test set is kept as the same as in LRS3. Finally, there are 41,560 utterances (1,500 speakers), 2,886 utterances (1,000 speakers) and 1,321 utterances (412 speakers) to simulate speech mixture in training, dev and test sets, respectively. 

\subsection{Data simulation}
To decouple visual cues, we simulate three kinds of dataset \footnote{\href{https://github.com/mrjunjieli/LRS3_for_AVSS}{https://github.com/mrjunjieli/LRS3\_for\_AVSS}}: different-speaker aligned-visual, different-speaker shuffled-visual and same-speaker aligned-visual. All speech mixtures are fully overlap. 

\textbf{different-speaker aligned-visual dataset:} two audios from different speakers are mixed between -5 $\sim$ 10 dB  in signal-to-interference ratio (SIR) \cite{vincent2006performance}. And the visual reference is time aligned visual sequence of target speaker. Finally, we simulate 41,558 (about 50 h), 2,884 (about 5 h) and 1,320 utterances for training, dev and test set, respectively. 

\textbf{different-speaker shuffled-visual dataset:} this is similar to the different-speaker aligned-visual dataset. Just the visual reference is shuffled each epoch during training. 

\textbf{same-speaker aligned-visual dataset:} two different utterances from the same speaker are mixed between -5 $\sim$ 10 dB  in signal-to-interference ratio (SIR). And the visual reference is time aligned visual sequence of target speaker. Finally, we simulate 41,338, 2,576 and 1,140 utterances for training, dev and test set, respectively.

\subsection{Training details}
We select Adam \cite{kingma2014adam} as an optimizer. The initial rate is set to $10^{-3}$. The learning rate is halved if the validation loss does not decrease for three epochs. The training process  stops when validation loss does not decrease consecutively for six epochs or training epoch reaches 100.

\section{Results}
\label{sec:result}

\subsection{Comparison with baselines}
Table \ref{tab:baseline} shows the performance of models under three simulated datasets. We evaluate the system's performance using SI-SNR and Perceptual Evaluation of Speech Quality (PESQ) \footnote{\href{https://github.com/vBaiCai/python-pesq}{https://github.com/vBaiCai/python-pesq}} \cite{rix2001perceptual}.

We evaluate the performance of AV-ConvTasNet-Sync and AV-ConvTasNet-Spk under `D-S S-V' dataset and `S-S A-V' dataset. AV-ConvTasNet-Sync and AV-ConvTasNet-Spk presents bad performance under `D-S S-V' and `S-S A-V', respectively, which indicates the effect of our proposed training strategies. AV-ConvTasNet-Sync only keeps synchronization cue and AV-ConvTasNet-Spk only keeps speaker identity cue. 

Since synchronization cue is not affected by speaker information, AV-ConvTasNet-Sync shows similar performance under different-gender mixtures and same-gender mixtures. The performance of AV-ConvTasNet-Spk shows speaker identity cue is also useful to perform separation, especially  under different-gender mixtures.   
These two models indicate that both two visual cues are useful for speaker extraction task, and synchronization cue is more important. The AV-ConvTasNet-Spk gets very bad performance when mixtures coming from the same gender, we guess that only using visual inputs are hard to distinguish  speakers, and it biases the optimization of model towards easy mixture examples.

By utilizing synchronization cue and speaker identity cue explicitly, the proposed DAVSE presents performance improvement over other models.  It proves the complementary effect of two visual cues compared to a single visual cue. And by modeling two visual cues explicitly, DAVSE also gets  higher evaluation results compared to AV-ConvTasNet. The results of `D-S S-V' and `S-S A-V' also show that when visual streams are out of synchronization, it gets a very poor performance, indicating the importance of synchronization cue.

We also observe that face input contains more information in terms of not only speaker identity cue but also synchronization cue. 
Previous works \cite{wang22s_interspeech, wu2019time, gao2021visualvoice} usually utilize lip streams to learn phonetic correlation between phoneme and lip motion.  Our results indicate that facial expressions contain more information in term of phonetic correlation.

\subsection{Visualization of visual embeddings}

\begin{figure}[t] 
\centering 
\includegraphics[width=0.47\textwidth]{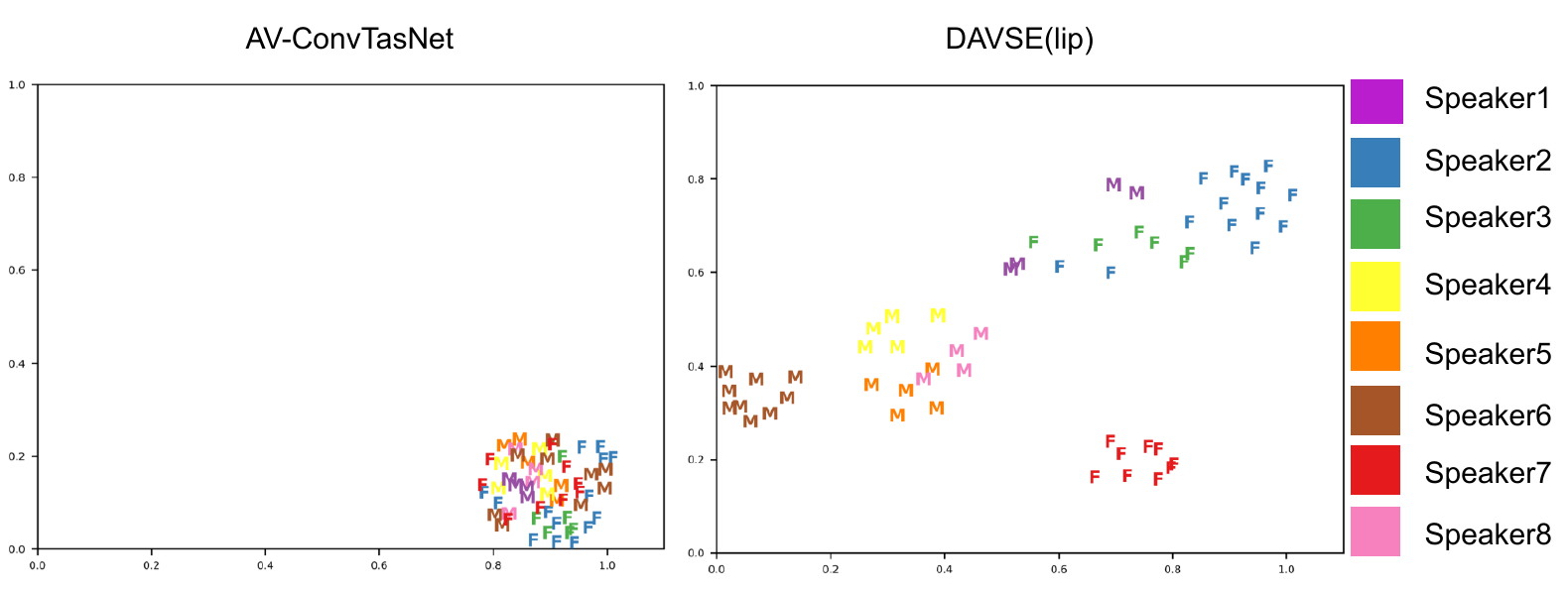}
\caption{The visual embeddings of 9 random speakers from test dataset visualized with UMAP \cite{mcinnes2018umap}. \textbf{M} and \textbf{F} denote the male and female, respectively. To compare different embeddings on the same scale, we choose to normalize them using min-max normalization, which scales them to a range between 0.0 $\sim$ 1.0}
\label{fig:visualization}
\end{figure}

 To visualize that DAVSE has learned a powerful visual embedding, Fig. \ref{fig:visualization} shows visual embeddings $V$ of 9 random speakers from AV-ConvTasNet and DAVSE using uniform manifold approximation and projection (UMAP). Compared to embeddings of AV-ConvTasNet, the DAVSE's  learned embeddings tend to distinguish not only speakers from different gender but also  speakers from same gender. Therefore, DAVSE is easier to extract target speech from its interfering speech. 

\section{Conclusions}
\label{sec: con}
In this work, we explore the role of visual cues in audio-visual speaker extraction. 
We propose two different training strategies to decouple the learning of
the synchronization and speaker identity cues. Experimental results show both visual cues are useful, while the  synchronization cue is at the higher end. We also propose a more explainable model, named
Decoupled Audio-Visual Speaker Extraction model (DAVSE), to
take advantage of both decoupled visual cues in speaker extraction.
Our DAVSE outperforms the baselines in terms of signal quality and
perceptual evaluations.


\section{Acknowledgements}
This work is supported by 1) Huawei Noah’s Ark Lab;
2) National Natural Science Foundation of China (Grant No. 62271432); 3) Guangdong Provincial Key Laboratory of Big Data Computing, The Chinese University of Hong Kong, Shenzhen (Grant No. B10120210117-KP02); 4) German Research Foundation (DFG) under Germany's Excellence Strategy (University Allowance, EXC 2077, University of Bremen); 5) Alibaba Innovative Research Program. 

\clearpage
\bibliographystyle{IEEEtran}
\bibliography{mybib}

\end{document}